\documentclass[letterpaper, 10 pt, conference]{ieeeconf}
\IEEEoverridecommandlockouts    
\usepackage[utf8]{inputenc}
\usepackage{amsmath}
\usepackage{amssymb}
\usepackage{graphicx}
\usepackage{babel}
\usepackage{balance}
\usepackage{cite}
\usepackage[position=bottom]{subfig}
\usepackage{comment}
\usepackage[T1]{fontenc}
\usepackage{mathtools}
    
\usepackage{amsmath}
\usepackage{amstext}
\usepackage{amssymb}
\usepackage{amsfonts}
\usepackage{float}

\usepackage{amsthm}  

\newtheorem{definition}{Definition}

\newtheorem{remark}{Remark}

\newtheorem{problem}{Problem}

\usepackage{algorithm}
\usepackage{algorithmicx} 
\usepackage{algpseudocode} %

\usepackage{todonotes}

\makeatletter
\let\NAT@parse\undefined
\makeatother

\usepackage{hyperref}
\hypersetup{
   colorlinks=True,
    linkcolor= blue,
    allcolors=blue,
    citecolor = blue,
    filecolor=black,      
    urlcolor=blue,
    }

\title{\LARGE \bf Re-Routing Strategy of Connected and Automated Vehicles\\ Considering Coordination at Intersections}

\author{Heeseung Bang, \textit{Student Member, IEEE}, Andreas A. Malikopoulos, \textit{Senior Member, IEEE}
    \thanks{This work was supported by NSF under Grants CNS-2149520 and CMMI-2219761.}
	\thanks{The authors are with the Department of Mechanical Engineering, University of Delaware, Newark, DE 19716, USA. (emails: \tt\small{heeseung@udel.edu}; \tt\small{andreas@udel.edu}.)}
}

\date{August 2022}

\begin{document}

\maketitle

\begin{abstract}
In this paper, we propose a re-routing strategy for connected and automated vehicles (CAVs), considering coordination and control of all the CAVs in the network. The objective for each CAV is to find the route that minimizes the total travel time of all CAVs. 
We coordinate CAVs at signal-free intersections to accurately predict the travel time for the routing problem.
While it is possible to find a system-optimal solution by comparing all the possible combinations of the routes, this may impose a computational burden. Thus, we instead find a person-by-person optimal solution to reduce computational time while still deriving a better solution than selfish routing. We validate our framework through simulations in a grid network.
\end{abstract}

\section{Introduction}

\PARstart{O}{ver} the last two decades, connected and automated vehicles (CAVs) have attracted considerable attention since they can improve safety, fuel economy, and other traffic conditions.
To reduce fuel consumption and travel time, several research efforts have focused on efficient coordination and control of CAVs at different traffic scenarios such as merging roadways \cite{Athans1969,Papageorgiou2002,Ntousakis:2016aa}, signal-free intersections \cite{Dresner2008,Au2015,Malikopoulos2020,chalaki2020TCST}, and corridors \cite{lee2013,Zhao2018ITSC,mahbub2020decentralized}.
These efforts have developed different coordination methods to reduce stop-and-go driving, which has resulted in shorter travel time with less energy use.

As another way to alleviate traffic issues, a significant number of research efforts have considered the routing problem of CAVs, which derived a solution for a single CAV and applied it to multiple vehicles.
Chen and Cassandras \cite{chen2020optimal} considered the dynamic vehicle assignment problem in a shared mobility system to optimizes the travel times of CAVs and the waiting times of passengers.
Tsao et al. \cite{tsao2019model} presented a similar approach for ride-sharing using model predictive control with fixed traffic conditions.
While these methods find the efficient route for a single CAV, they are not verified for multiple CAVs applications.
Some other approaches proposed in the literature include graph search algorithms \cite{sachenbacher2011efficient}, reinforcement learning techniques \cite{liang2020mobility,wan2018model,gammelli2021graph}, and mixed-integer linear programs \cite{boewing2020vehicle,chen2019integrated,salazar2019optimal}.
Furthermore, other research efforts considered variation of basic formulations by including different types of vehicles \cite{huang2020eco,salazar2019optimal}, adding battery constraints \cite{pourazarm2014optimal}, and considering interaction with infrastructures \cite{rossi2019interaction,estandia2021interaction}.

In addition to a single CAV, many studies have also focused on larger-scale interactions between multiple CAVs.
In the majority of these studies, a travel latency function was used to estimate travel delay caused by other CAVs \cite{houshmand2019penetration,salazar2019congestion,wollenstein2020congestion}.
Recent research efforts have developed a method for routing and relocating CAVs in more complex networks \cite{wollenstein2021routing} and considered situations where charging scheduling of electric CAVs is required \cite{bang2021AEMoD}.
However, none of these efforts considered the impact of microscopic phenomena such as coordination and control of CAVs on traffic conditions.

To the best of our knowledge, there have been only a few studies focusing on the routing problem with consideration of interactions between CAVs and their actual movements.
Chu et al. \cite{chu2017dynamic} solved the routing problem for each CAV considering traffic by employing a dynamic lane reversal approach. However, they simply approximated the travel time of CAVs proportional to the number of CAVs on each road segment.
Likewise, Mostafizi et al. \cite{mostafizi2021decentralized} developed a decentralized routing framework with a heuristic algorithm, but they estimated travel speeds inversely proportional to the number of CAVs on the roads.
In our previous work \cite{Bang2022combined}, we introduced a new routing framework combined with the coordination of CAVs.
the framework predicts traffic conditions from coordination information and finds the optimal route for each new travel request.
However, we have imposed an assumption that a new CAV does not affect previously planned trajectories.
This assumption sometimes causes the situation where a CAV entering the network has no solution because it yields all the roads that it can travel through.
In addition, all CAVs only find their route at the beginning of the trip because, once the trajectory is fixed, it does not get affected by any other CAVs' decisions, which is quite a strong assumption to apply in reality.

In this paper, we propose a re-routing framework for CAVs with consideration of traffic conditions resulting from each CAV's decision.
The framework is built upon our previous work \cite{Bang2022combined}, which consists of two parts: (i) at an intersection level, we formulate a coordination problem to predict travel time and acquire an exact trajectory, and (ii) we formulate a routing problem using the predicted travel time at a network level.
To avoid computational challenge, we generate some candidate routes for each CAV and find the best combination of the routes to reduce the complexity of the problem and make it possible to apply this framework to more extensive networks.
The main contribution of this work is to advance the state-of-the-art congestion-aware routing problem in a way that makes the problem more realistic by relaxing those assumptions.
Another contribution is that we provide a method that always yields a person-by-person optimal solution, which is a series of the best choice for each CAV.

The remainder of this paper is structured as follows. In Section \ref{sec:problem}, we formulate the routing problem and coordination problem. In Section \ref{sec:routing}, we present algorithms to solve the problems and discuss the optimality of the solution. We validate our method through the simulations results in Section \ref{sec:simulation}. Finally, in Section \ref{sec:conclusion}, we conclude and discuss some directions for future work.

\section{Problem Statement} \label{sec:problem}

In this paper, we develop a framework that consists of multiple optimization problems at two different levels. The upper level contains a routing framework for a road network. We consider having a finite number of candidate routes and formulate optimization problems to find the best combination of those routes. At the low level, we coordinate CAVs to cross intersections fast without violating any safety constraints, and a travel time computed from coordination is directly involved in the upper-level optimization as a cost function.
The detailed problem formulation will be explained next.

\subsection{Trips and Routing on Road Network}

\begin{figure}
    \centering
    \includegraphics[width=0.85\linewidth]{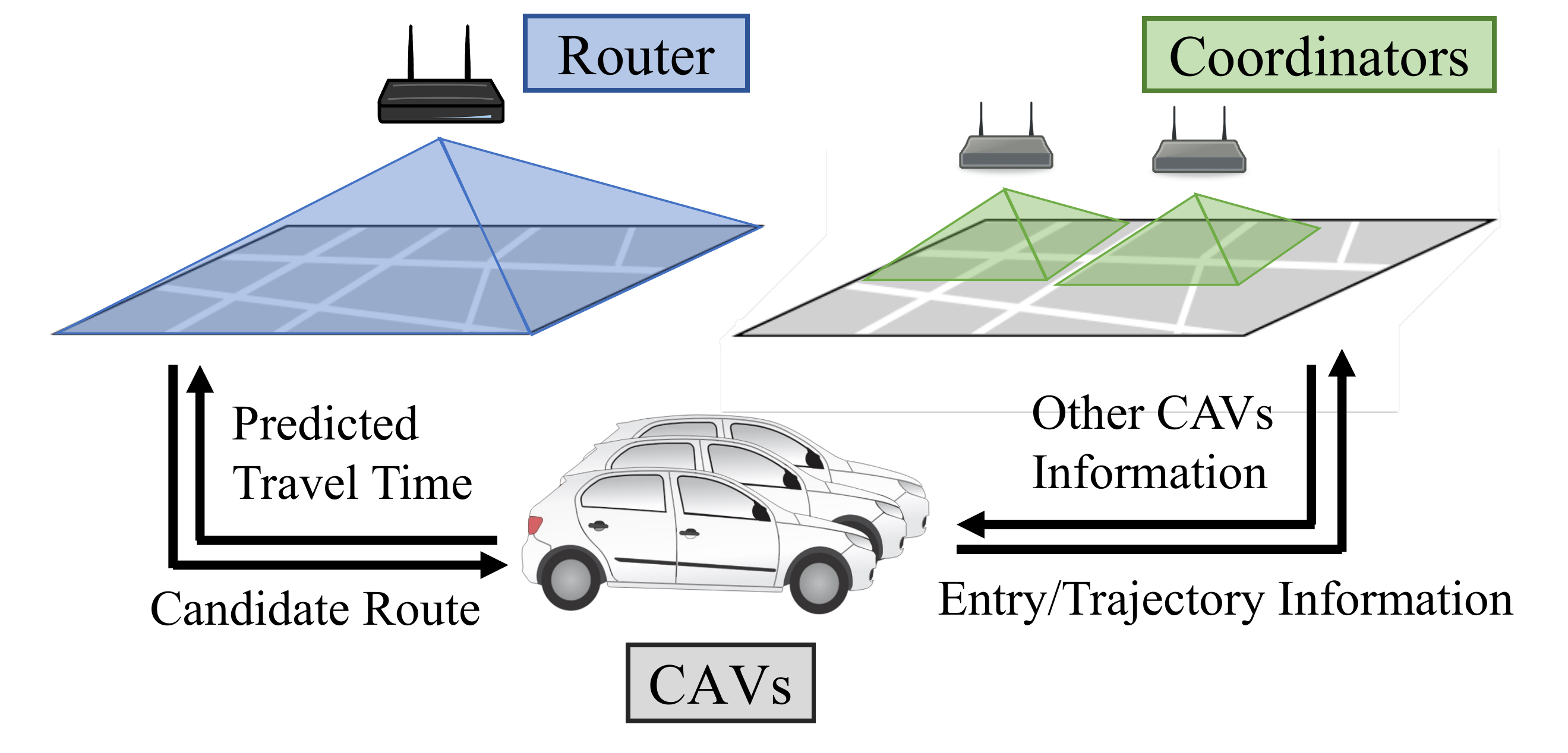}
    \caption{The concept diagram of the routing process.}
    \label{fig:concept}
\end{figure}

We consider a road network given by a directed graph $\mathcal{G} = (\mathcal{V}, \mathcal{E})$, where $\mathcal{V}$ is a set of nodes and $\mathcal{E}$ is a set of edges (roads). Each node is a junction point of two different intersections (exiting from one intersection and entering another one), and each node can simultaneously function as the origin for one trip and the destination for another trip.
In the network, there exists a \textit{router} that communicates with CAVs and selects the best combination of routes and a \textit{coordinator} for each intersection which shares the geometry of intersections and trajectories of other CAVs (Fig. \ref{fig:concept}).

Our framework considers a $100\%$ penetration rate of CAVs in the road network. Let $\mathcal{N}(t) = \{1,\dots,N(t)\}$ be a set of CAVs traveling in the network at time $t\in\mathbb{R}_{\geq0}$, where $N(t)\in\mathbb{N}$ is the total number of CAVs at moment.
For each CAV $i\in\mathcal{N}(t)$, we have trip information $\mathbb{I}_i =(o_i,d_i,t_i^\mathrm{s})$ which consists of origin $o_i\in\mathcal{V}$, destination $d_i\in\mathcal{V}$, and the starting time of travel $t_i^\mathrm{s} \leq t$.
This implies that we receive trip information only when a new CAV starts a trip and do not have any information about trips departing in the near future.
We assume that each CAV joins the network and departs from $o_i$ at time $t_i^\mathrm{s}$ and that there are enough CAVs at the origin nodes.
To relax this assumption, one can consider a whole trip of CAVs departing from and returning to stations \cite{Bang2022combined} or relocation of empty CAVs \cite{bang2021AEMoD}.

In our previous work \cite{Bang2022combined}, we assumed that a newly introduced CAV does not affect previous CAVs' trajectories.
This assumption made it possible to solve a substantial centralized problem in a computationally distributed manner.
However, this is a strong assumption in reality and also yields cases where no solution exists when a new CAV is introduced.
To relax this assumption, we let the router to re-route all the CAVs in the network within new candidate routes whenever a new CAV joins the network and finds a system-optimal solution.

At every time $t$, when a new CAV joins the network, we generate $M\in\mathbb{N}$ different candidate routes for each CAV $i\in\mathcal{N}(t)$, which connect the current location of CAV $i$ to the destination $d_i$. We denote $\mathcal{M} = \{1,\dots, M\}$ to be an index set of those routes and let $\mathcal{P}_i = \{\mathcal{P}_{i,1},\dots,\mathcal{P}_{i,M}\}$ denote a set of all candidate routes for the CAV $i$, where $\mathcal{P}_{i,m} \subset \mathcal{E}$ is a $m$-th candidate route of the CAV $i$.
This paper assumes that a finite number of candidate routes are determined by a high-level decision maker and given to the router.
Next, we define an \textit{assignment matrix} to match the CAVs to one of their candidate routes.

\begin{definition} \label{def:assignment}
    Let \textit{assignment matrix} $\mathbf{A}(t)$ be a binary matrix that maps all CAVs $i\in\mathcal{N}(t)$ to a unique route $\mathcal{P}_{i,m}$ for $m\in\mathcal{M}$.
\end{definition}
Note that $\mathbf{A}(t)=[\mathbf{A}_1,\dots,\mathbf{A}_{N(t)}]$, where $\mathbf{A}_i = [a_{i,1},\dots,a_{i,M}]^\mathsf{T}$ is an \textit{assignment vector} for CAV $i\in\mathcal{N}(t)$. 
The assignment matrix can be determined by solving an optimization problem, which we define next.

\begin{problem}[Optimal Routing] \label{prb:global_routing}
    We find a system-optimal combination of the routes by solving the following problem
    \begin{align}\label{eq:opt_route}
        \min_{\mathbf{A}} ~& \Big\{ \sum_{i\in\mathcal{N}(t)} \sum_{m\in\mathcal{M}} a_{i,m} T_{i}(\mathbf{A}(t),m) \Big\}\\
        \emph{subject~to:}& \sum_{m\in\mathcal{M}} a_{i,m} = 1,\quad \forall i \in\mathcal{N}(t), \nonumber \\
        & a_{i,m} \in \{0,1\},\quad\forall  i\in\mathcal{N}(t), m\in\mathcal{M}, \nonumber
    \end{align}
    where $T_{i}(\mathbf{A}(t),m)$ is the travel time of CAV $i$ following the $m$-th candidate route.
    The constraints ensure each CAV to be assigned to a unique route.
\end{problem}
The travel time $T_{i}(\mathbf{A}(t),m)$ can be computed using the coordination framework, which will be presented later in this section.

\subsection{Coordination at Intersections}

\begin{figure}
    \centering
    \includegraphics[width=0.7\linewidth]{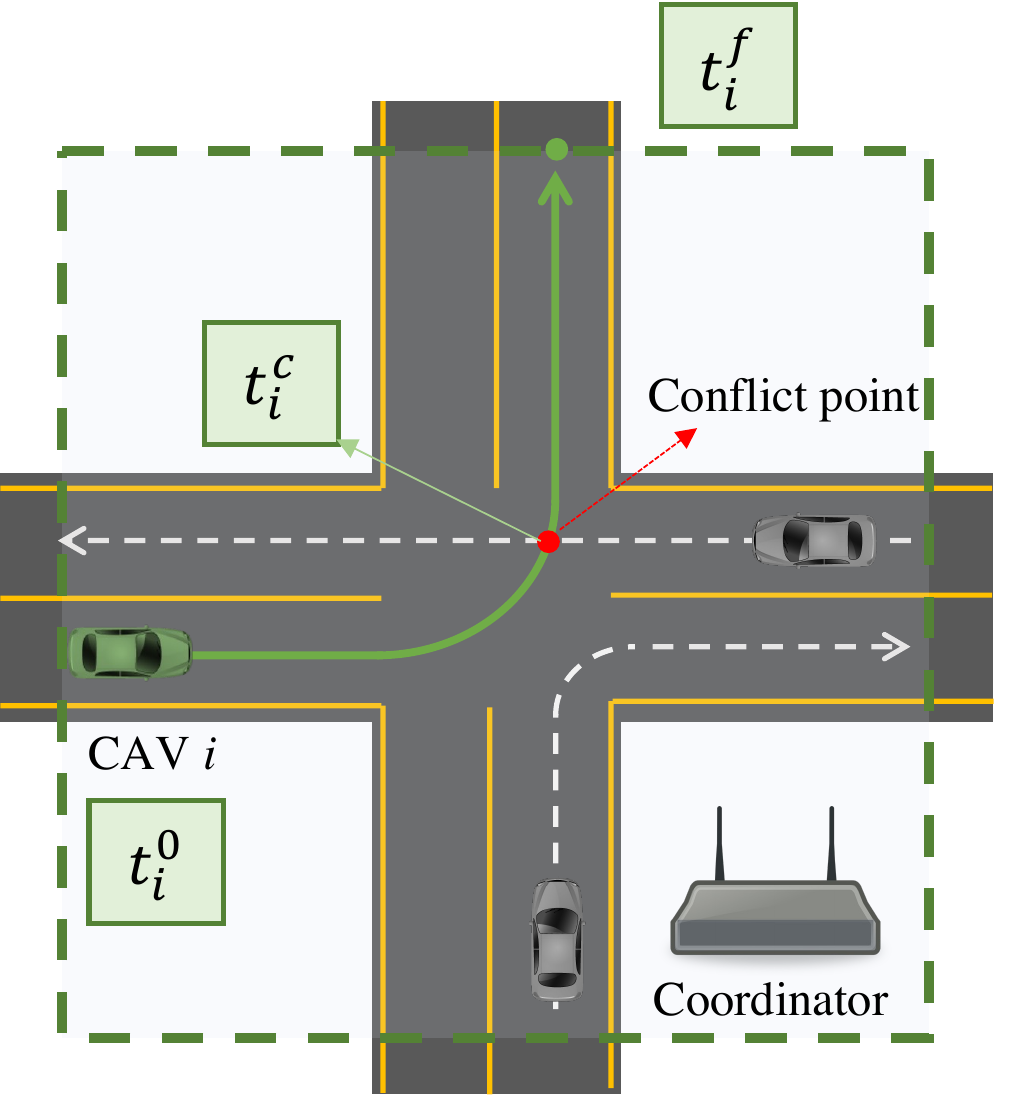}
    \caption{Coordination of CAV $i$ at an intersection. CAV $i$ enters at $t_i^0$, passes the conflict point at $t_i^c$, and exits at $t_i^f$.}
    \label{fig:coordination}
\end{figure}

Given the routes of all the trips, we now model intersections and coordinate CAVs to travel without any collision.
Let $\mathcal{L}=\{1,\dots,L\}$ denote a set of signal-free intersections, where $L\in\mathbb{N}$ is the total number of intersections.
We let $\mathcal{N}_l(t) \subset \mathcal{N}(t)$ denote a set, which includes all the CAVs in the intersection $l\in\mathcal{L}$.
Each intersection consists of four entry nodes, four exit nodes, and twelve edges connecting each entry node to the exit nodes of the other three directions (Fig. \ref{fig:coordination}). 
The geographical junction points of the edges are called \textit{conflict points}, where lateral collisions may occur, and we define an index set of conflict points denoted by $\mathcal{C}\subset\mathbb{N}$.
For each intersection, there exists a coordinator that communicates with CAVs to share all the geographical information and trajectories of other CAVs in the intersection.
We assume that a coordinator can communicate with CAVs without any delays or errors, and each CAV plans its trajectory in the order of entrance to the intersection.

Next, we determine a model and constraints of CAVs.
Each CAV $i\in\mathcal{N}(t)$ follows doubled integrator dynamics
\begin{equation}\label{eqn:dynamics}
    \begin{aligned}
    \dot{p}_i(t) &= v_i(t),\\
    \dot{v}_i(t) &= u_i(t),
    \end{aligned}
\end{equation}
where $p_i(t)\in\mathbb{R}$, $v_i(t)\in\mathbb{R}$, and $u_i(t)\in\mathbb{R}$ are position, speed, and control input at time $t$, respectively.
The state and input constraints are given as
\begin{align}
    u_{i,\text{min}} & \leq u_i(t) \leq u_{i,\text{max}}, \label{eqn:ulim}\\
    0 < v_{\text{min}} & \leq v_i(t) \leq v_{\text{max}}, \label{eqn:vlim}
\end{align}
where $u_{i,\text{min}}$, $u_{i,\text{max}}$ are the lower bound and the upper bound of control inputs, which are the hardware limits on acceleration, and $v_{\text{min}}$, $v_{\text{max}}$ are the minimum and maximum speed limits, respectively.

To find CAVs' trajectories at an intersection, we consider a coordination framework similar to the one introduced in \cite{Malikopoulos2020}.
The main idea of the framework is to derive energy-optimal trajectories for CAVs and find proper parameters that do not violate any constraint.
Given dynamics \eqref{eqn:dynamics}, we adopt an energy-optimal unconstrained trajectory for CAVs, which minimizes acceleration/deceleration.
The trajectory for each CAV $i\in\mathcal{N}_l(t)$ is derived as
\begin{align} \label{eq:optimalTrajectory}
    u_i(t) &= 6 a_i t + 2 b_i, \notag \\
    v_i(t) &= 3 a_i t^2 + 2 b_i t + c_i, \\
    p_i(t) &= a_i t^3 + b_i t^2 + c_i t + d_i, \notag
\end{align}
where $a_i, b_i, c_i$, and $d_i$ are constants of integration.
We determine these constants using the boundary conditions
\begin{align}
     p_i(t_i^{0}) &= 0,\quad  v_i(t_i^0)= v_i^0 , \label{eq:bci}\\
     p_i(t_i^f)&=p_i^f,\quad u_i(t_i^f)=0, \label{eq:bcf}
\end{align}
where $t_i^0$ and $t_i^f$ are the entry and exit time of CAV $i$ on the intersection $l$. 
Note that the final speed $v_i(t_i^f)$ varies with respect to $t_i^f$, hence, by convention we consider that $u_i(t_i^f)=0$ \cite{bryson1975applied}.
For the detailed derivation of the energy-optimal trajectory, see \cite{Malikopoulos2020}.

Since the trajectories in \eqref{eq:optimalTrajectory} are derived from an unconstrained optimization problem, there is no safety guarantee for using the trajectories.
Thus, we find exit time $t_i^f$ that satisfies all the safety and constraints so that even unconstrained trajectories can avoid any collision.
In terms of safety constraints, we only consider the collision between the CAVs in the same intersection $l\in\mathcal{L}$.
We impose the following constraints to avoid a rear-end collision between a CAV $i\in\mathcal{N}_l(t)$ and a preceding CAV $k\in\mathcal{N}_l(t)$ in the same road,
\begin{equation}
    p_k(t)-p_i(t) \geq \delta_i(t) = \rho + \varphi \cdot v_i(t).  \label{eqn:rearend1}
\end{equation}
Here, $\delta_i(t)$ is the safety distance for CAV $i$, which depends on the standstill distance $\rho\in\mathbb{R}$, reaction time $\varphi\in\mathbb{R}$, and the speed of CAV $i$.
Next, we consider two different scenarios for lateral safety at a conflict point.
Suppose there exists a CAV $k\in\mathcal{N}_l(t)$ with the already planned trajectory that shares a conflict point with a CAV $i$.
Then, CAV $i$ can pass the conflict point either earlier or later than CAV $k$.
In the case of CAV $i$ passing the conflict point later than CAV $k$, the trajectory of CAV $i$ must satisfy
\begin{equation}
    p_i^c - p_i(t) \geq \delta_i(t),\quad\forall t \in [t_i^0,t_k^c], \label{eqn:lateral1}
\end{equation}
where $p_i^c\in\mathbb{R}$ is the location of the conflict point $c\in\mathcal{C}$ on CAV $i$'s path, and $t_k^c$ is the known time that CAV $k$ reaches at the conflict point $c\in\mathcal{C}$.
On the other hand, if CAV $i$ passes the conflict point before CAV $k$, the constraint becomes
\begin{equation}
    p_k^c - p_k(t) \geq \delta_k(t) = \rho+\varphi\cdot v_k(t),\quad\forall t\in[t_k^0,t_i^c], \label{eqn:lateral2}
\end{equation}
where $p_k^c\in\mathbb{R}$ is the location of the conflict point on CAV $k$'s path, and $t_i^c$ is the time when CAV $i$ reaches the conflict point $c\in\mathcal{C}$, which is determined by the trajectory of CAV $i$.
Since we use \eqref{eq:optimalTrajectory} with a constraint \eqref{eqn:vlim}, the inverse of position always exists, i.e., $t_i(\cdot)=p_i^{-1}(\cdot)$.
We call this a \textit{time trajectory} of CAV $i$ \cite{Malikopoulos2020}, which yields an arrival time at a certain point.
For example, we obtain the time $t_i^c = p_i^{-1}(p_i^c)$ at which CAV $i$ arrives at the conflict point $c\in\mathcal{C}$ along the energy optimal trajectory \eqref{eq:optimalTrajectory}.

To guarantee lateral safety, either \eqref{eqn:lateral1} or \eqref{eqn:lateral2} must be satisfied.
Thus, we impose the lateral safety constraint by taking the minimum of \eqref{eqn:lateral1} and \eqref{eqn:lateral2}, i.e.,
\begin{align}\label{eq:lateralMinSafety}
    \min \Bigg\{ &\max_{t\in[t_i^0, t_k^c]} \{ \delta_i(t) + p_i(t) - p_i^c\}, \notag\\
            &\max_{t\in[t_k^0, t_i^c]} \{ \delta_k(t) + p_k(t) - p_k^c \}   \Bigg\} \leq 0. 
\end{align}

To reduce travel time of CAV $i$, we find the minimum exit time $t_i^f$.
We define the feasible set $\mathcal{T}_i=\left[\underline{t}_i^{f}, \overline{t}_i^f\right]$, where $\underline{t}_i^{f}$ and $\overline{t}_i^f$ are the earliest and latest exit time that CAV $i$ can possibly have, considering all the limits \eqref{eqn:ulim}, \eqref{eqn:vlim} and boundary conditions \eqref{eq:bci}, \eqref{eq:bcf} \cite{chalaki2020experimental}.

\begin{problem}
To find minimum exit time, each CAV $i\in\mathcal{N}_l(t)$ solves the following optimization problem
    \begin{align}\label{eq:tif}
        &\min_{t_i^f\in \mathcal{T}_i} t_i^f \\
        \emph{subject to: }&
         \eqref{eq:optimalTrajectory} - \eqref{eqn:rearend1}, \eqref{eq:lateralMinSafety}\notag.
    \end{align}
    \label{pb:timeMinRP}
\end{problem}

The solution to Problem \ref{pb:timeMinRP} becomes the entry time of the next intersection or the final travel time if CAV $i$ reached the final destination $d_i$.

\section{Person-by-Person Optimal Re-Routing} \label{sec:routing}

In this section, we present a solution approach and analyze an optimality of the solution.
In general, if a new CAV joins the system for a new trip, the previous routes may not be optimal anymore.
Thus, the router solves Problem \ref{prb:global_routing} to obtain the system-optimal solution whenever there is a change in the trips.
However, computational cost for solving Problem \ref{prb:global_routing} gets tremendously larger as the number of CAVs $N(t)$ increases, because it requires to compare all $T_{i}(\mathbf{A}(t),m)$ values for all possible $\mathbf{A}(t)$.
As the assignment matrix $\mathbf{A}(t)$ can have $M^{N(t)}$ different cases, the increase in the number of CAVs will exponentially scale the computational cost.
Therefore, rather than solving the global optimization problem for all CAVs, we modify it to a single CAV routing problem.

\begin{problem}[Single CAV Routing] \label{prb:pbp_routing}
    For a single CAV $i\in\mathcal{N}(t)$, we find its optimal route by solving the following problem
    \begin{align}\label{eq:route}
        \min_{a_{m}\in\mathbf{A}_i} ~& \Big\{ \sum_{m\in\mathcal{M}} a_{m} T_{i}(\mathbf{A}(t),m)+ \sum_{j\in\mathcal{N}(t)\setminus\{i\}} T_{j}(\mathbf{A}(t),\bar{m}_j) \Big\}\\
        \emph{subject}& \emph{ to:} \sum_{m\in\mathcal{M}} a_{m} = 1, \nonumber \\
        & \emph{~~~~~} a_{m} \in \{0,1\},\quad\forall m\in\mathcal{M}. \nonumber
    \end{align}
    Here, $\mathbf{A}_i$ is an assignment vector for CAV $i$, and $\bar{m}_j$ is a pre-selected route for CAV $j\in\mathcal{N}(t)\setminus\{i\}$.
\end{problem}
In this problem, all the other CAVs' routes are fixed, and only CAV $i$ selects its route, which minimizes the total travel time of all the CAVs.

\begin{algorithm}
 \caption{Event-Triggered Routing Process}
 \begin{flushleft}
        \textbf{Input:} $t$, $\mathbb{I}_{N(t)}$, $\mathbf{A}=[\mathbf{A}_1,\dots,\mathbf{A}_{N(t)-1}]$\\
        \textbf{Output:} {$\mathbf{A}^*(t)$}
\end{flushleft}

\begin{algorithmic}[1]
\State{Update location of each CAV $i\in\mathcal{N}(t)$}
\State{Generate new route $\mathcal{P}(t) := \{\mathcal{P}_{i,m}\big|\forall i \in\mathcal{N}(t), m\in\mathcal{M}\}$}
\For{$m\in\mathcal{M}$}
    \State{$\tau_m^\mathrm{tmp} \gets T_{N(t)}(\mathbf{A}(t),m)$}
\EndFor
\State{$m^* \gets \arg\min_m \tau_m^\mathrm{tmp}$}
\State{$\tau^* \gets \tau_{m^*}^\mathrm{tmp}$}
\State{$\mathbf{A}^*(t) \gets [\mathbf{A}, \mathbf{A}_{N(t)}]$} \Comment{Add new assignment vector}
\State{$\mathbf{A}^*(t) \gets$\texttt{Re-Routing}$(\mathbf{A}^*(t),\tau^*,\mathcal{P}(t))$}
\end{algorithmic} \label{Alg:Routing}
\end{algorithm}

\begin{algorithm}
 \caption{Re-Routing Process}
 \begin{flushleft}
        \textbf{Input:} {$\mathbf{A}^*(t), \tau^*, \mathcal{P}(t)$}\\
        \textbf{Output:} {$\mathbf{A}^*(t)$}
\end{flushleft}

\begin{algorithmic}[1]
\State{$q \gets N(t)$} \Comment{Check the last modified CAV}
\While{\texttt{true}}
    \For{$i\in\mathcal{N}(t)$}
        \State{$\tau_m^\mathrm{tmp} \gets T_{i}(\mathbf{A}^*(t),m),~~\forall m\in\mathcal{M}$}
        \State{$m^* \gets \arg\min_m \tau_m^\mathrm{tmp}$}
        \State{$\Tilde{\tau} \gets \tau_{m^*}^\mathrm{tmp}$}
        \If{$\Tilde{\tau} < \tau^*$}
            \State{$\tau^* \gets \Tilde{\tau}$}
            \If{$a_{i,m^*} \neq 1$} \Comment{$i$ changed the route}
                \State{$q \gets i$}
                \State{Update $\mathbf{A}^*_i$ with route $\mathcal{P}_{i,m}$}
            \EndIf
        \EndIf
        \If{$q=i$} \Comment{No one changed the route}
            \State{\texttt{Return} $\mathbf{A}^*$;}
        \EndIf
    \EndFor
\EndWhile
\end{algorithmic} \label{Alg:reRouting}
\end{algorithm}

As Problem \ref{prb:pbp_routing} deals with a single CAV, we solve the problem whenever a new CAV joins the network.
Algorithm \ref{Alg:Routing} explains the process of solving Problem \ref{prb:pbp_routing} for new CAV $N(t)$.
When CAV $N(t)$ joins the network, the router receives candidate routes for all CAVs in the network.
Then, the router communicates with CAV $N(t)$ and selects a route that minimizes total travel time.
Although it tries to minimize the increase in total travel time, CAV $N(t)$'s new route would naturally affect other CAVs' trajectories and delay their trips.
Therefore, we also need to check whether the change of other CAVs' trajectories would reduce the travel time or not and re-select their routes if it is beneficial.
The re-routing process is presented in Algorithm \ref{Alg:reRouting}.
The router fixes all the CAVs' route except one CAV and finds the route with minimum total travel time.
It repeats the process for all CAVs until none of them has the benefit of changing the route.

The assignment $\mathbf{A}^*(t)$ resulted from Algorithm \ref{Alg:reRouting} is \textit{person-by-person} optimal as none of CAVs can improve the travel time by changing its own decision\cite{Malikopoulos2021}.
\begin{remark} \label{rem:convergence}
    Note that CAVs have a finite number of candidate routes and the total travel time has a lower bound, which is the summation of each CAV's travel time on empty roads.
    Since the router modifies the combination of routes only if there is a better combination, the routes will be fixed in a finite number of comparison.
\end{remark}

\begin{algorithm}[t]
 \caption{Prediction of Total Travel Time for All CAVs}
 \begin{flushleft}
        \textbf{Input:} {$\mathbf{A}(t), i, m$}\\
        \textbf{Output:} {$\tau^{tot}$}
\end{flushleft}

\begin{algorithmic}[1]
\State{Assign CAV $i$ to $m$-th route}
\State{Define initial speed of all CAVs}
\State{$\tau_n \gets$ CAV $n$'s arrival time at a nearest node $\forall n\in\mathcal{N}(t)$}
\State{$\tau^{tot} \gets 0$}
\While{\texttt{CAVs traveling}}
    \State{$n^* \gets \arg\min_n \tau_n$}
    \State{$t_{n^*}^0 \gets \tau_{n^*}$}
    \State{Find intersection $l$ satisfying $n^*\in\mathcal{N}_l(t)$}
    \State{$t_{n^*}^f \gets$ Solution to Problem \ref{pb:timeMinRP}}
    
    \State{Update CAV $n^*$'s corresponding intersection}
    \State{$\tau_{n^*} \gets t_{n^*}^f$}
    
    \If{CAV $n^*$ reached $d_{n^*}$}
        \State{$\tau^{tot} \gets \tau^{tot} + \tau_{n^*}$}
        \State{Mark $n^*$ as \texttt{done}}
    \EndIf
\EndWhile
\end{algorithmic} \label{Alg:coordination}
\end{algorithm}

Given $N(t)$ CAVs with $M$ routes, we need to compare $M^{N(t)}$ different cases to get system-optimal solution (solution to Problem \ref{prb:global_routing}).
Meanwhile, the re-routing process in Algorithm \ref{Alg:Routing} stops comparing if it reaches the person-by-person optimal solution.
In the worst-case scenario, Algorithm \ref{Alg:Routing} will compare all $M^{N(t)}$ cases and yield a system-optimal solution.
\begin{remark} \label{rem:computation}
    Algorithm \ref{Alg:Routing} would yield either a person-by-person optimal solution with lower computational cost or a system-optimal solution with the same computational cost with solving Problem \ref{prb:global_routing}.
    Comparing more different cases during the re-routing process means that it is getting closer to the system-optimal solution.
\end{remark}

Next, we present Algorithm \ref{Alg:coordination}, which explains the method of computing the total travel time of all the CAVs.
Considering CAV $i$ selected $m$-th route, we update the assignment matrix.
At a certain time $t$, all the CAVs are located either on the edges or the nodes.
The CAVs on the edges already have the trajectories of their current intersection, which allow us to know when they will arrive at the next nodes.
Considering CAVs on the nodes, we coordinate those CAVs at their located intersections in the order of arrival time on the nodes.
By solving Problem \ref{pb:timeMinRP}, we get the arrival time to the next node, and we repeat this process until all the CAVs reach their destinations.

\section{Simulations}   \label{sec:simulation}

This section provides the simulation results of CAVs traveling in a road network.
Through the simulations, we investigate the benefit of our method:
(i) computational benefit of acquiring person-by-person optimal solution and (ii) combining the routing process with coordination of CAVs.

\subsection{Simulation Setup}

We considered a grid network ($25$ intersections, $71$ nodes, and $170$ edges) and  generated $100$ random trips.
Whenever a new CAV starts the trip, the router generates three different routes ($M=3$) that are shortest-time paths based on (1) empty roads, (2) statistical traffic data, and (3) traffic conditions at the moment.
For the baseline scenario, we assume that each CAV selects the shortest-time path at the moment of departure.
To evaluate the effectiveness of combining the routing process with coordination of CAVs, for both the proposed method and the baseline scenario, we let CAVs communicate with the coordinator at each intersection and plan their trajectories to pass the intersection without any collision.
The difference is in the route selection mechanism, where the proposed method uses a prediction of the future traffic condition while the baseline scenario considers the traffic condition at the moment.

\begin{figure}[t]
    \centering
    \includegraphics[width=0.9\linewidth]{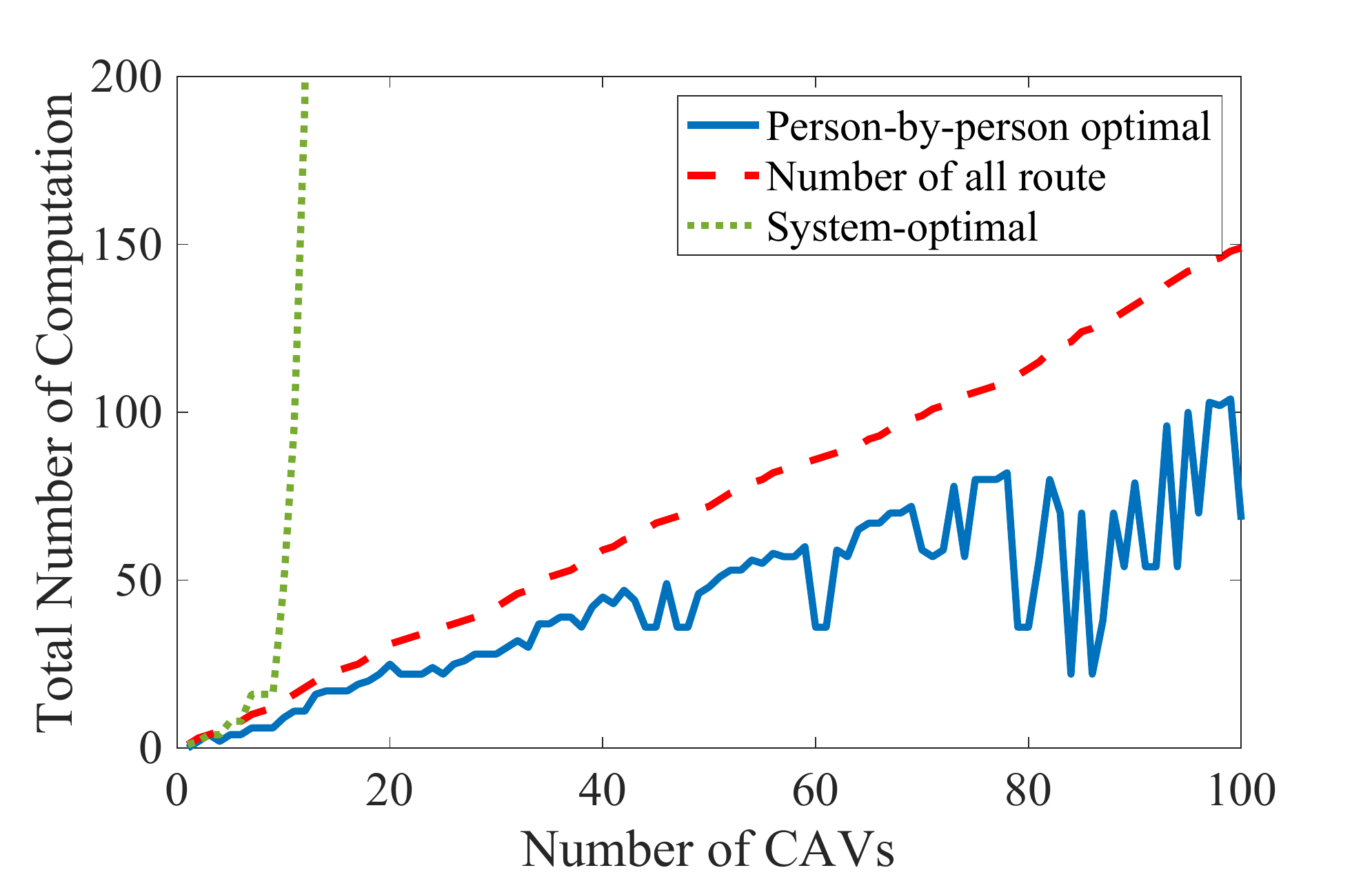}
    \caption{The number of computations for a system-optimal approach (dotted-green) and a person-by-person optimal approach (solid-blue). The total number of candidate routes is shown in a dashed-red line.}
    \label{fig:case}
\end{figure}
\begin{figure}[t]
    \centering
    \includegraphics[width=0.9\linewidth]{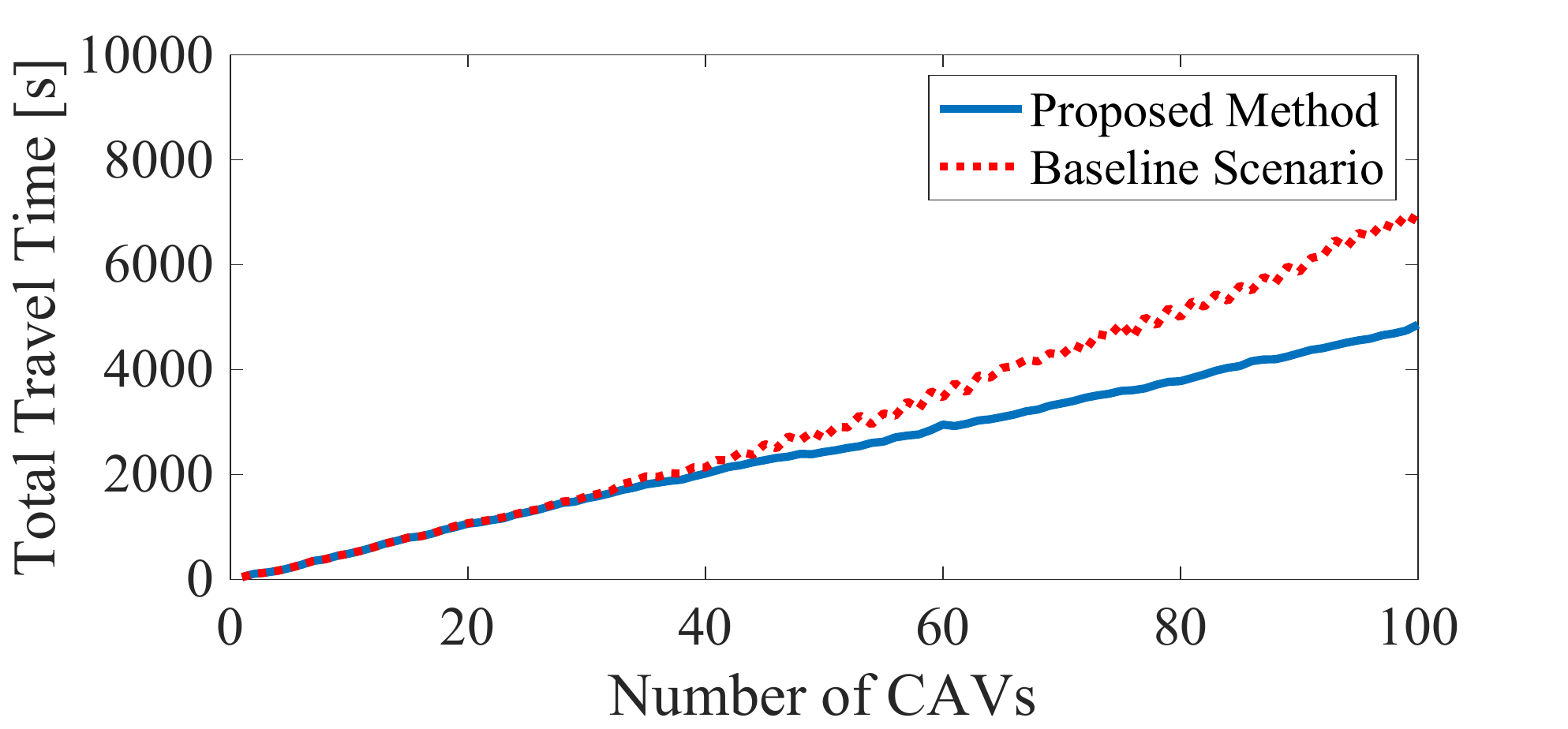}
    \caption{Total travel time of the proposed method (blue) and the baseline scenario (red).}
    \label{fig:traveltime}
\end{figure}

\subsection{Simulation Results}

First, we compared the number of computations in Fig. \ref{fig:case} to evaluate the computational cost of different algorithms.
To reduce the computational cost, we searched person-by-person optimal solutions only among the CAVs that were delayed by other CAVs.
This method made it possible to find a person-by-person optimal solution without checking all CAVs resulting in less computation than the number of all routes.
The number of computations for Problem \ref{prb:global_routing} increased up to $2.98\times 10^{25}$ while the proposed method required $68$ computations.
Figure \ref{fig:traveltime} illustrates the total travel time of CAVs increasing as the number of CAVs increased.
In the baseline scenario, new CAVs naturally delayed other CAVs' trips, even though they selected the shortest-time path considering the traffic condition at the moment.
On the other hand, by using coordination information, the proposed method found the route with smaller delays.

\section{Concluding Remarks} \label{sec:conclusion}

In this paper, we proposed a re-routing strategy for CAVs using coordination and control information. We formulated an optimal route selection problem with candidate routes for all CAVs. The problem was solved in a distributed manner, which resulted in yielding a person-by-person optimal solution. We demonstrated the effectiveness of our method through the simulations. 
Future research should consider generating optimal candidate routes for each CAVs and analyzing performance bounds of the person-by-person optimal solution.


\bibliographystyle{IEEEtran}
\bibliography{Bang, IDS_02182023, SharedMobilityRef,TAC_references}

\end{document}